\documentclass[%
 reprint,
superscriptaddress,
 amsmath,amssymb,
 aps,
]{revtex4-1}

\usepackage{graphicx}
\usepackage{dcolumn}
\usepackage{bm}

\usepackage[normalem]{ulem} 
\usepackage{color} 
\begin{document}

\preprint{APS/123-QED}

\title{Towards a universal law for blood flow}

\author{Alexander Farutin}
    \affiliation{Laboratoire Interdisciplinaire de Physique, Universit\'{e} Grenoble Alpes and CNRS, F-38000 Grenoble, France}
\author{Abdessamad Nait-Ouhra}
    \affiliation{Laboratoire Interdisciplinaire de Physique, Universit\'{e} Grenoble Alpes and CNRS, F-38000 Grenoble, France}
    \affiliation{LaMCScI, Faculty of Sciences, Mohammed V University of Rabat, 1014 Morocco}
    \author{Mehdi Abbasi}
    \affiliation{Laboratoire Interdisciplinaire de Physique, Universit\'{e} Grenoble Alpes and CNRS, F-38000 Grenoble, France}
    \author{Gopal Dixit}
    \affiliation{Laboratoire Interdisciplinaire de Physique, Universit\'{e} Grenoble Alpes and CNRS, F-38000 Grenoble, France}    \author{Hamid Ez-zahraouy}     \affiliation{LaMCScI, Faculty of Sciences, Mohammed V University of Rabat, 1014 Morocco}       
    \author{Othmane Aouane}
    \affiliation{Helmholtz Institut (HIERN)
Forschungszentrum J\"ulich GmbH
F\"urther Stra{\ss}e  248
90429 N\"urnberg, Germany}  \author{Jens Harting}
  \affiliation{Helmholtz Institut (HIERN)
Forschungszentrum J\"ulich GmbH
F\"urther Stra{\ss}e  248
90429 N\"urnberg, Germany}  \author{Chaouqi Misbah}
    \email{chaouqi.misbah@univ-grenoble-alpes.fr}
    \affiliation{Laboratoire Interdisciplinaire de Physique, Universit\'{e} Grenoble Alpes and CNRS, F-38000 Grenoble, France}

\date{\today}

\begin{abstract}
Despite decades of research on blood flow, an analogue of Navier-Stokes equations that accurately describe blood flow properties has not been established yet. The reason behind this is that the properties of blood flow seem \`a priori  non universal as they depend on various factors such as global concentration of red blood cells (RBCs) and channel width. Here, we have discovered a universal law when the stress and strain rate  are measured at a given  local RBCs concentration. However, the local concentration must be determined in order to close the problem. We  propose a non-local diffusion equation of RBCs concentration that agrees with the full simulation.  The universal law is exemplified for both shear and pressure driven flows. While the theory is restricted to a simplistic geometry (straight channel)  it provides a fundamental basis for future research on blood flow dynamics and could lead to the development of a new theory that accurately describes blood flow properties under various conditions, such as in complex vascular networks. 

\end{abstract}

\pacs{Valid PACS appear here}
\maketitle


\paragraph*{Introduction-}
Almost two centuries ago, Poiseuille \cite{Poiseuille1836} reported that the distribution of red blood cells (RBCs) in animal (frog) blood vessels is non-uniform, with a cell-free layer near the vessel wall and an RBC-rich zone toward the center. A century later, Fahraeus and Lindqvist \cite{faahraeus1931viscosity} conducted {\it in vitro} experiments on blood flow in a tube subject to a pressure difference and observed that the cell-free layer has a significant effect on the apparent viscosity: viscosity decreases with increasing confinement. 

Experimental\cite{Chien1960,Lipowsky2005} and direct simulation studies\cite{Zhang2008RedBC,Aidun,Bagchi,Kruger,Ishikawa,Liu} on blood (taken as a suspension of RBCs, modeled as elastic capsules) have so far focused on the global properties of the flow (such as the apparent viscosity). The global properties depend on several parameters (like channel width, global RBC concentration), without any simple rational analytical expression. In addition, information on local flow properties (like flow patterns) are desirable in many circumstances.
The derivation of a local constitutive law, such as the one relating stress to local shear rate, would therefore  be a considerable advance, opening the way to predictive studies. 




The objective of this Letter is to explore, for simplified geometry (straight channel) and simple system (without taking into account the aggregation process among RBCs), the possibility for the existence of a universal law. The crux of the study is the identification of the appropriate variables that enter the constitutive law. We demonstrate that when the stress and strain rates are evaluated for a given local concentration of RBCs in the channel, the constitutive law acquires a universal character that is independent of  the global concentration of RBCs and of the channel width. This universal law predicts a relation between the local stress and the local strain rate and local fluidity (or viscosity) which is a function of the local concentration of RBCs. 
It will be seen  that the extracted law is valid  in both linear shear and pressure-driven flows.

Determining the local concentration of RBCs is a key challenge in developing a closed form of the constitutive law. To address this issue, we propose a  model that predicts the concentration profile. Two antagonist effects compete in fixing the RBCs concentration profile: (i) the lift migration of RBCs due to the walls, which pushes RBCs towards the channel center, and (ii) hydrodynamic interactions among RBCs, which spreads RBCs throughout the channel, known as hydrodynamic diffusion. Our analysis reveals that the concentration profile obeys a nonlocal diffusion equation. Combined with the local stress-strain rate relation with a local fluidity function of local concentration, this leads to a complete set of constitutive laws describing blood flow in the channel.

\paragraph*{Model and methods-}
We consider two models to study cell suspensions: a suspension of vesicles in 2D and a suspension of capsules (a well-adopted model for RBCs) in 3D. These suspensions are subject to shear flow with shear stress $\sigma_0$ and a pressure-driven flow with an imposed pressure gradient $\Delta p/L$, where $L$ is the channel length.

The 2D cells are modeled as inextensible fluid membranes, while the 3D cells are represented by elastic capsules. Both systems have resistance to bending, which is characterized by the bending energy $E=(\kappa/2) \int H^2 dA$, where $\kappa$ is the bending modulus of the cell, $H$ is the mean curvature, and $dA$ is the arclength in 2D or area in 3D. In 3D, the membrane has an additional elastic energy given by $\mu_s(I_1^2 + 2I_1-I_2)/12+ \kappa_\alpha I_2^2/12$, where $\mu_s$ is the shear elastic modulus, $\kappa_\alpha$ is the area dilation modulus, and $I_1$ and $I_2$ are the in-plane strain invariants (see \cite{Krueger2011}).

The fluids inside and outside the cells are taken in 2D to have the same viscosity $\eta$, and with viscosity ratio equal to 5 in 3D. The suspensions are characterized by the global areal (2D) and volume (3D) concentration, denoted by $\phi_0$ in both cases. The cell size $R_0$ is defined as $R_0=\sqrt{A/\pi}$ in 2D and $R_0=[3V/(4\pi)]^{1/3}$ in 3D, where $A$ is the enclosed area (2D) and $V$ is the enclosed volume (3D).

An important geometric parameter of the cell is its reduced volume in 3D $\nu\equiv (R_0/\sqrt{A/(4\pi)})^3$ (reduced area $\nu\equiv (2\pi R_0/(p))^2$ in 2D), where $A$ and $p$ are the surface area and the perimeter of the cell in 3D and 2D, respectively. For reference, we take $\kappa \simeq 3 ; 10^{-19};$ J and $\mu_s\simeq 4 $ $\mu$ N/m, which are values known for red blood cells \cite{Suresh2006}. The extension modulus is taken to be large enough to prevent local extension of the membrane. In 2D the capillary number is $Ca = \mu \dot{\gamma}_w R_0^3 / \kappa$, and in 3D $Ca = \mu \dot{\gamma}_w R_0/ \mu_s, $where $\dot{\gamma}_w$ is the wall shear rate.
For a healthy RBC, we have $R_0\simeq 2.7 \mu m$, and $\nu$ is reported to be in the range $\simeq 0.6-0.64$ \cite{linderkamp83,Fung2013}. Here, we choose $\nu=0.65$ as a reference \cite{Li2013,cordasco14,Lanotte:2016}.

The cells are arranged in 2D  within a straight channel that extends for a length $L$, with periodic boundary conditions enforced along the flow direction (with length $L_x$ along $x$). The width of the channel is denoted as $W$, and we impose no-slip boundary conditions at the walls of the channel. In 3D, we consider periodic boundary condition along $x$ (flow direction) and $z$ directions (with dimensions $L_x$ and $L_z$). The suspension is confined in the $y$ direction. In 2D, we employ a boundary integral formulation, and in 3D we use a lattice Boltzmann method. Further details on the numerical methods used can be found in Refs. \cite{Thiebaud2014, Aouane17, Krueger2011}.

\paragraph*{Results-}
At the particle scale, the suspension is strongly heterogeneous due to discontinuities in parameters such as the local stress $\sigma(\mathbf r,t)$ and concentration $\phi(\mathbf r,t)$. To define the local concentration $\phi(\mathbf r,t)$, we set its value to 1 if the position vector $\mathbf r$ is inside a cell and 0 if it is outside. Our approach involves averaging local quantities along the flow direction (and the vorticity direction in 3D), where the problem is translationally invariant. Additionally, we use time-averaging after skipping the initial transient to compensate for the finite size of the computational domain. This yields the average values as a function of the lateral position $y$. We mark the $x$-averaged ($x,z$-averaged in 3D) values with a bar above the corresponding symbol, while values averaged over the entire channel (e.g., global RBC concentration) are marked with a superscript 0. Of utmost importance are the average local concentration $\bar\phi(y)$, average local shear stress $\bar\sigma_{xy}(y)$, and average local strain rate $\bar{\dot{\gamma}}(y) = \partial_y \bar{u}_x(y)$.

\begin{figure}
\begin{center}
\includegraphics[height=4 cm,width=0.99\columnwidth]{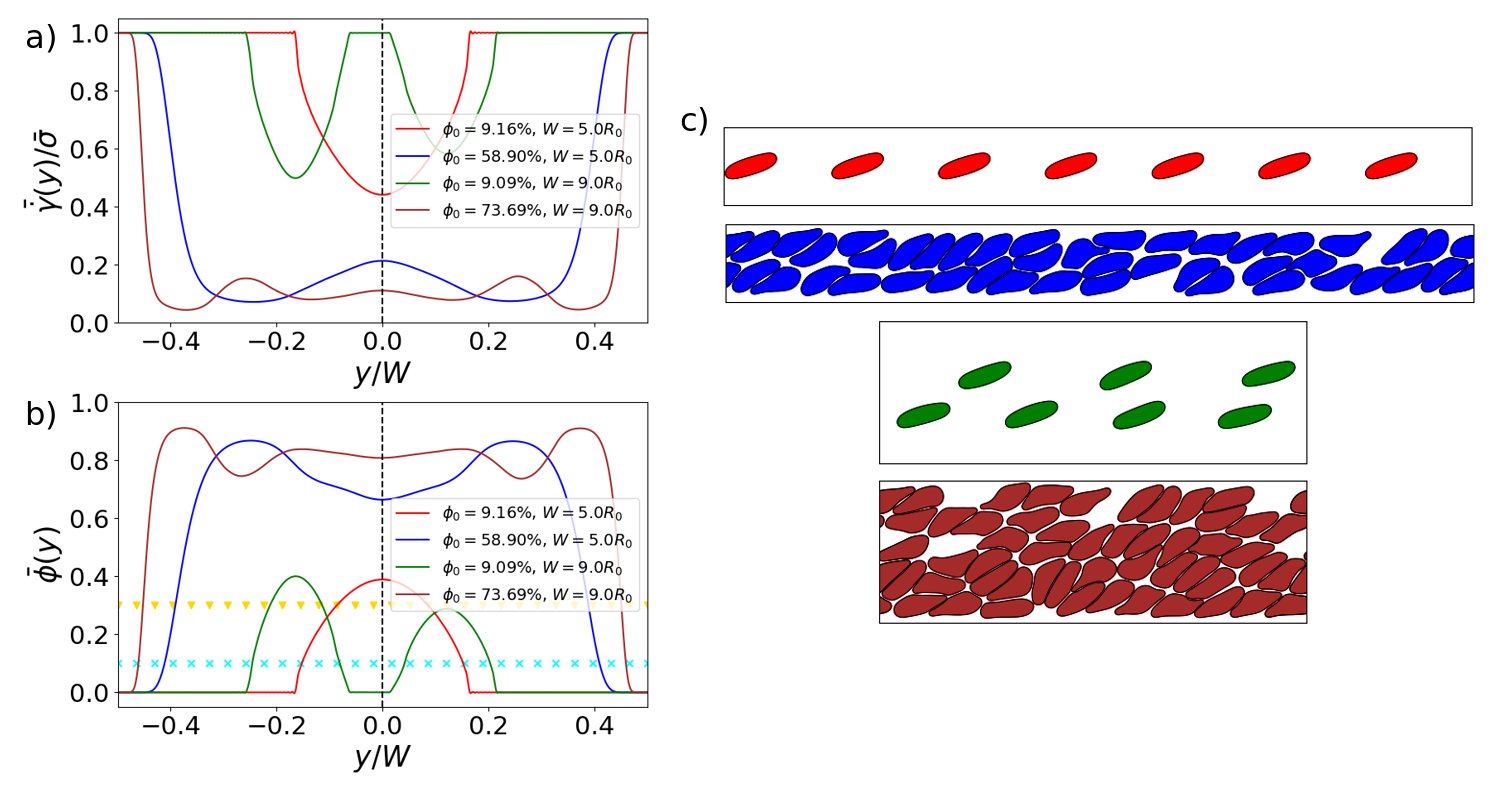}
\caption{\label{figshear} (a): Fluidity as a function of the imposed stress for several values of global concentration and channel width. (b): local concentration $\bar{\phi (y)} $. (c): cell spatial configurations; we used same colors for the corresponding fluidity. The symbols in (b) are explained below in text. }
\end{center}
\end{figure}

In 2D, we have conducted an extensive exploration of various parameters, including $W$ ($W \in [5R_0, 17R_0]$) and the average concentration of the suspension $\phi_0$ ($\phi_0 \in [0.05, 0.75]$). We first examined the case of an imposed linear shear flow, $\dot{\gamma_0}$, which is achieved by moving the bounding walls in opposite directions. By applying momentum conservation principles alone (without invoking any constitutive law of the suspension), we can show that the stress $\bar{\sigma}_{xy}(y)$ remains constant within the gap (see SI) and is denoted as $\bar{\sigma}$. Figure \ref{figshear}a displays the ratio $\bar{\dot{\gamma}}(y)/\bar{\sigma}$ as a function of $y$, for different global concentrations $\phi_0$ and channel widths $W$. This ratio corresponds to the inverse of the viscosity (or fluidity), which is constant for a simple fluid. However, in the case of the suspension, this quantity exhibits significant variations within the gap, and depends on both $\phi_0$ and $W$.

\begin{figure}
\begin{center}
\includegraphics[width=0.9\columnwidth]{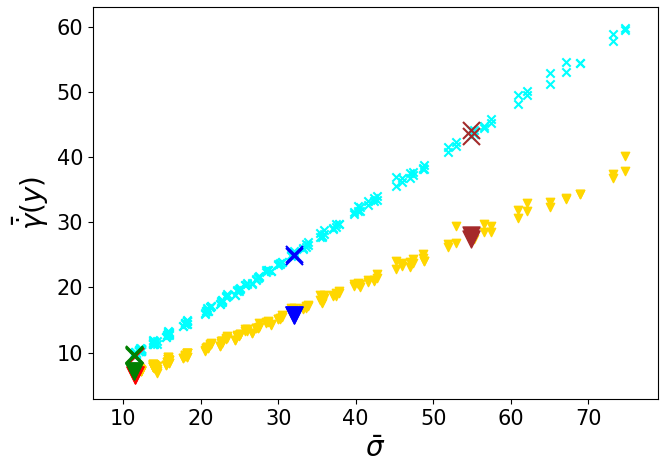}
\caption{\label{profile} The stress-strain rate relation measured at a given local concentration (corresponding to the 2 examples of local concentrations given in Fig.\ref{figshear}b). The 4 crosses and 4 triangles in blue, red, green and brown  correspond to the 4 intersections of the two  horizontal lines (representing the two selected local concentrations) with the local concentration concentration profiles of the same colors in Fig.\ref{figshear}b. The numerous light blue and yellow symbols correspond to data obtained for different global concentrations and channel widths providing different local concentration profiles and thus different location of intersection with the two fixed local concentrations (crosses and triangles in Fig.\ref{figshear}}
\end{center}
\end{figure}

The  oscillations of $\bar{\dot{\gamma}}(y)/\bar{\sigma}$ in Fig. \ref{figshear}a are correlated with  the local concentration profiles $\bar{\phi}(y)$ (Fig. \ref{figshear}b; see also the spatial cellular organization in Fig. \ref{figshear}c).  In this representation, we did not find any  rescaling of the data that would lead to a collapse in Fig. \ref{figshear} for different $\phi_0$ and $W$. This raises the question of whether there is a hidden universality, and if so, how to uncover it.
\paragraph*{Universal law}:
It is found that representing data not at a given position $y$, but at a given local concentration $\bar\phi(y)$ sheds light on the intricate nature of the behavior of the suspension, allowing us to reveal a universal law.  We fix the value of a local concentration (indicated, for example, by crosses in Figure \ref{figshear}b),  and determine the corresponding values of $y$ for which  the stress and the strain rate are measured. This procedure is repeated  (still for the same local concentration) for all concentration profiles obtained for different global concentrations $\phi_0$ and channel widths $W$.  Then all obtained couples of strain rate and stress are reported in the plane ($\bar{\dot{\gamma}}(y), \bar{\sigma}$).
Importantly, we observed that data for the same local concentration (crosses in Figure \ref{figshear}b) collapsed onto the same line in Figure \ref{profile}. We then select another local concentration   (for example triangles in Figure \ref{figshear}b)) and repeat the same procedure. Here again we observe data collapse  (Figure \ref{profile}). Note the slopes in Figure \ref{profile} depend very weakly on shear stress (a very weak shear-thinning; see SI), for which no special concern is needed here.


\begin{figure}
\begin{center}
\includegraphics[width=0.48\columnwidth]{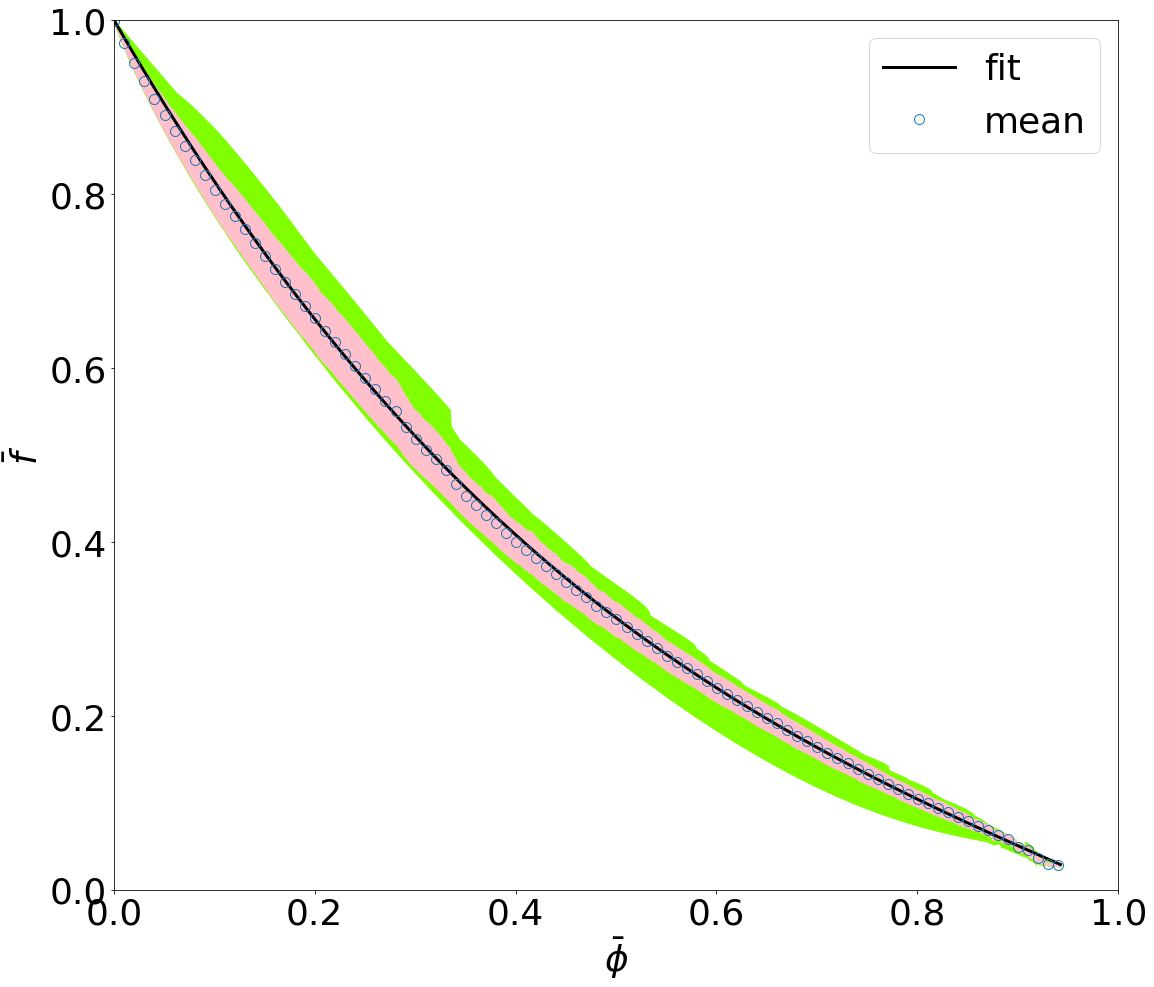}
\includegraphics[width=0.48\columnwidth]{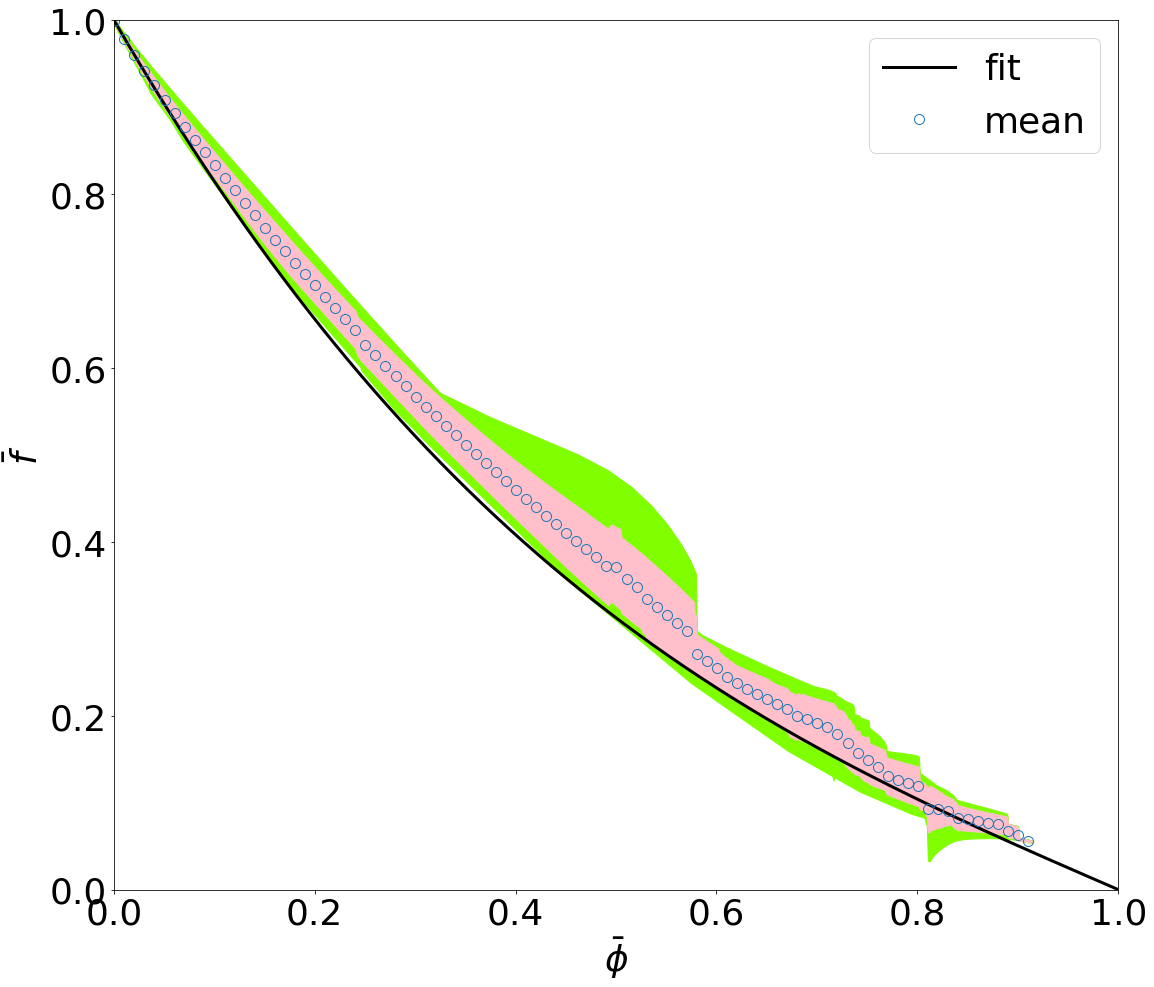}\caption{\label{fit}Local rheological laws for all explored channel widths and concentrations in 2D shear flow (left) and Poiseuille flow (right). The green zone shows the maximum and minimum values observed in all simulations. The pink zone shows the mean dispersion of the results (mean$\pm$m.s.d.). The symbols are the mean values and the curve is the fit with eq. (\ref{fit_law}).}
\end{center}
\end{figure}

To obtain a more comprehensive understanding, we repeated this procedure for all available local concentrations (instead of only the two illustrated ones as done in Figure \ref{figshear}b) and for the entire parameter range explored ($W \in [5R_0, 17R_0]$ and $\phi_0 \in [0.05, 0.75]$). Similar behavior for the stress-strain rate relation as in Figure \ref{profile} was observed, with each line corresponding to a given local concentration having a different slope. As shown in Figure \ref{fit}, all the slopes obtaiend for each fixed local concentration, $\bar{\dot{\gamma}}(y)/\bar{\sigma}$ (called local fluidity), were gathered together as functions of the local concentration, highlighting a universal behavior in which local fluidity depends only on local concentration and not on channel width $W$ or global concentration $\phi_0$. From these results, we conclude that the local rheological law has the form $\bar{\sigma}_{xy}(y) = \bar{\eta}(y) \bar{\dot{\gamma}}(y)$, or equivalently,
\begin{equation}
\label{fluidity}
\bar{\dot{\gamma}}(y) = \bar{f}(y) \bar{\sigma}_{xy}(y),
\end{equation}
where $\bar{\eta}(y)$ and $\bar{f}(y)$ are the local viscosity and fluidity, respectively.

Figure \ref{fit} presents the mean viscosity obtained by averaging the local viscosity for a given concentration over all available simulations. The figure shows that the averaged law effectively describes all available data, as indicated by the mean standard deviation interval and the maximum and minimum values. Notably, the mean standard deviation remains below 0.04 for all local concentrations. However, some outlier conditions exhibit a discrepancy between the actual local law and the averaged one as large as 0.11. These cases typically correspond to ordered states in which the suspension breaks into a set of files (see Fig.\ref{figshear}c). Such states are mostly observed at low concentrations or in narrow channels.

To fit the dependence of local fluidity on local concentration, we used a simple expression that agrees well with the numerical results, as demonstrated in Figure \ref{fit}. The expression is given by:
\begin{equation}
\label{fit_law}
\bar f=(1-\bar\phi)\left[1+(1-\bar\phi)^2\right]/2.
\end{equation}

We conducted further systematic simulations of pressure-driven flow, specifically Poiseuille flow, and found that the same universal law (\ref{fluidity}) with the same fluidity (\ref{fit_law}) applies. Figure \ref{fit} (right) shows the fluidity as a function of local concentration for all parameter sets. We observed similar qualitative behavior in both shear and Poiseuille flow; however, we noted that eq. (\ref{fit_law}) slightly underestimates the fluidity in Poiseuille flow.

\begin{figure}
\begin{center}
\includegraphics[width=0.9\columnwidth]{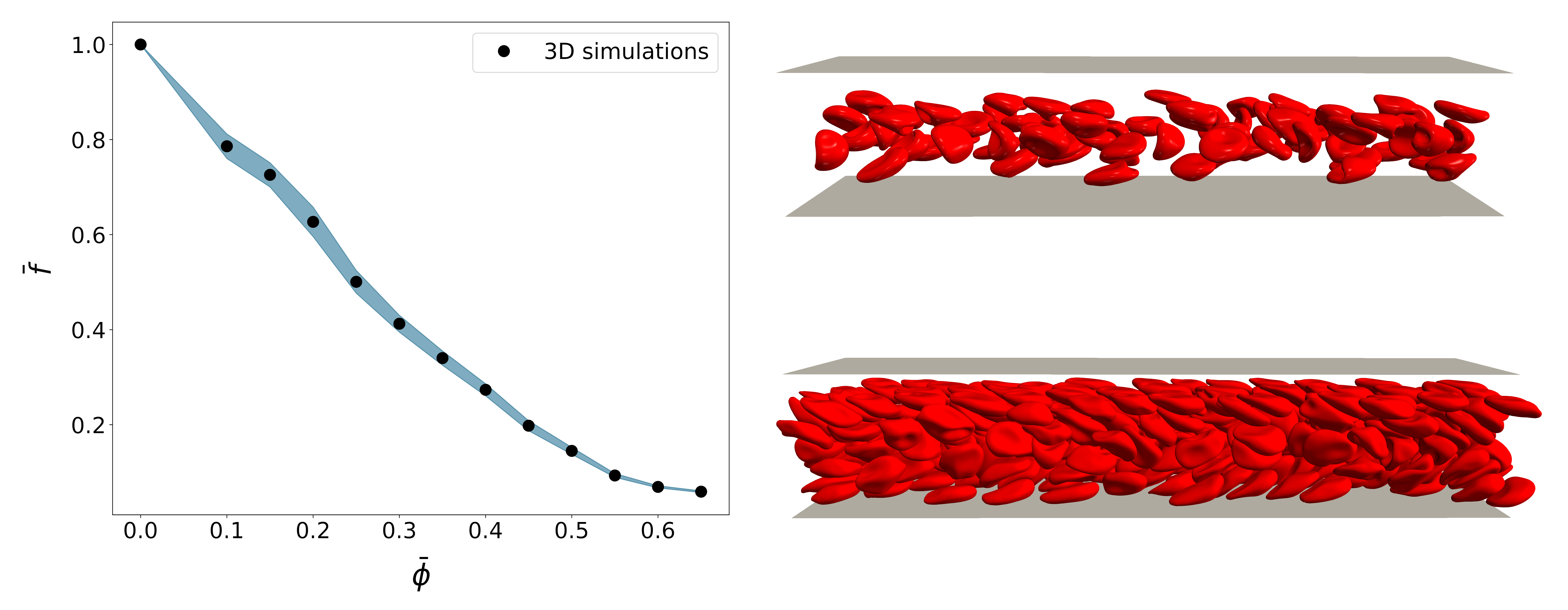}
\caption{\label{3Df} Left: Fluidity as a function of local concentration (symbols correspond to average, and color to mean square deviation). Right: RBCs configuration for $\phi_0=0.1$ and $\phi_0=0.4$.}
\end{center}
\end{figure}

 \paragraph*{3d results}  The RBCs are modeled as biconcave capsules with a reduced volume $\nu = 0.65$. The viscosity contrast between the inner and outer fluids is fixed to 5 (a typical value for human RBCs).  We have performed simulations for channel widths $L_y=5R_0$, $7R_0$ and $9R_0$. The global volume fraction is varied from $\phi_0=0.1$ to $\phi_0=0.5$. The capillary number is fixed to $Ca=1.6$ for all the simulations (the saturated regime). 
  Following the same approach as in 2D, we averaged over the flow direction (i.e., along $x$) and the vorticity direction (i.e., along $z$) and confirmed the validity of the universal law (\ref{fluidity}). The fluidity as a function of local concentration $\bar \phi (y)$ is shown in Fig. \ref{3Df}, highlighting data collapse.


\paragraph*{Determination of local concentration}
The local concentration $\bar\phi(y)$ is unknown a priori and must be determined in order to solve the problem at hand. Two factors contribute to the concentration profile $\bar\phi(y)$: (i) the wall-induced lift \cite{cantat1999lift,coupier2008noninertial}, 
 and (ii) the hydrodynamic diffusion\cite{eckstein1977,leighton1987,Davies,Tan2012Hydrodynamic} of the cells, where hydrodynamic interactions tend to spread the cells uniformly within the suspension. The balance between these two effects defines the stationary concentration profile in the channel.

The conservation law for particle concentration reads
\begin{equation}
\label{flux}
\partial_t\bar\phi+\partial_y\bar J_y=0,  \;\bar J_y=\bar J^l_y+\bar J^r_y,
\end{equation}
where $\bar J_y(y)$ is the lateral flux of average local concentration, $\bar J^l$ is the lift-induced flux and $\bar J^R$ accounts for particle-particle interactions.
The stationary profile corresponds to $\bar J_y=0$, where the cell spreading due to hydrodynamic diffusion is balanced by the lift that tends to focus the suspension at the center of the channel.

The lift-induced flux $\bar J_{cy}^l$ of the cell centers can be expressed as\cite{Farutinlaws}\begin{equation}
\label{Jl}
\bar J_{cy}^l(y)=\bar\phi_c(y)L(y), \; L(y)=\frac{L_0\sigma_0y}{(W/2-R_0)^2-y^2},
\end{equation}
where $L(y)$ is the lift function and $L_0$ is a numerical constant related to cell properties (e.g. the reduced area).

The hydrodynamic diffusion current is classically described by a Fick-like law\cite{eckstein1977,leighton1987,Davies,Tan2012Hydrodynamic}
$\bar J_y^r=-D(\bar\phi)\partial_y\bar\phi$
where $D(\bar\phi)$ is the diffusion coefficient, with $D\sim \dot\gamma \phi$. 
However, this law cannot accurately describe the profile shown in Fig.\ref{figshear}b: the concentration gradient $\partial_y\bar\phi$ is zero at a concentration extrema
implying that $\bar J_y=\bar J^l_y$ at such points.
The lift flux $\bar J^l_y$ is only zero at the center of the channel and is directed towards the center otherwise for the conditions analyzed here. Thus, there is a non-zero particle flux (at extrema outside the center of the channel in Fig.\ref{figshear}b), which contradicts the stationary distribution assumption. A non local diffusion  resolves this dilemma. 
The mass current  (see SI) contains a first term which re uncorrelated fluctuations from different files, resulting in a random walk of the cell in the $y$ direction. The second term is a  deterministic repulsion leading to a lateral velocity that is related to the concentration profile $\bar\phi_c$. This is expressed as:
\begin{equation}
\label{Jr}
\bar J_{cy}^r=-D(y)\partial_y\bar\phi_c(y)+\bar\phi_c(y)\int\partial_y U(y-y')\bar\phi_c(y')dy',
\end{equation}
where $\bar\phi_c(y)$ represents the concentration of center of masses of cells; $D$ and $U$ are given in SI. Note that nonlocal diffusion was also presented earlier in the context of blood flow\cite{Shaqfehnonlocal,Grahamnonlocal}, but the present formulation is relatively simpler.
Overall, we found that the non-local migration law reproduces well the main qualitative features of the concentration profiles (as those presented in Fig.\ref{figshear}b) for a diverse set of $\phi_0$ and $W$ and can even give quantitatively correct predictions in many cases (see SI). 

\begin{figure}
\begin{center}
\includegraphics[width=0.8\columnwidth]{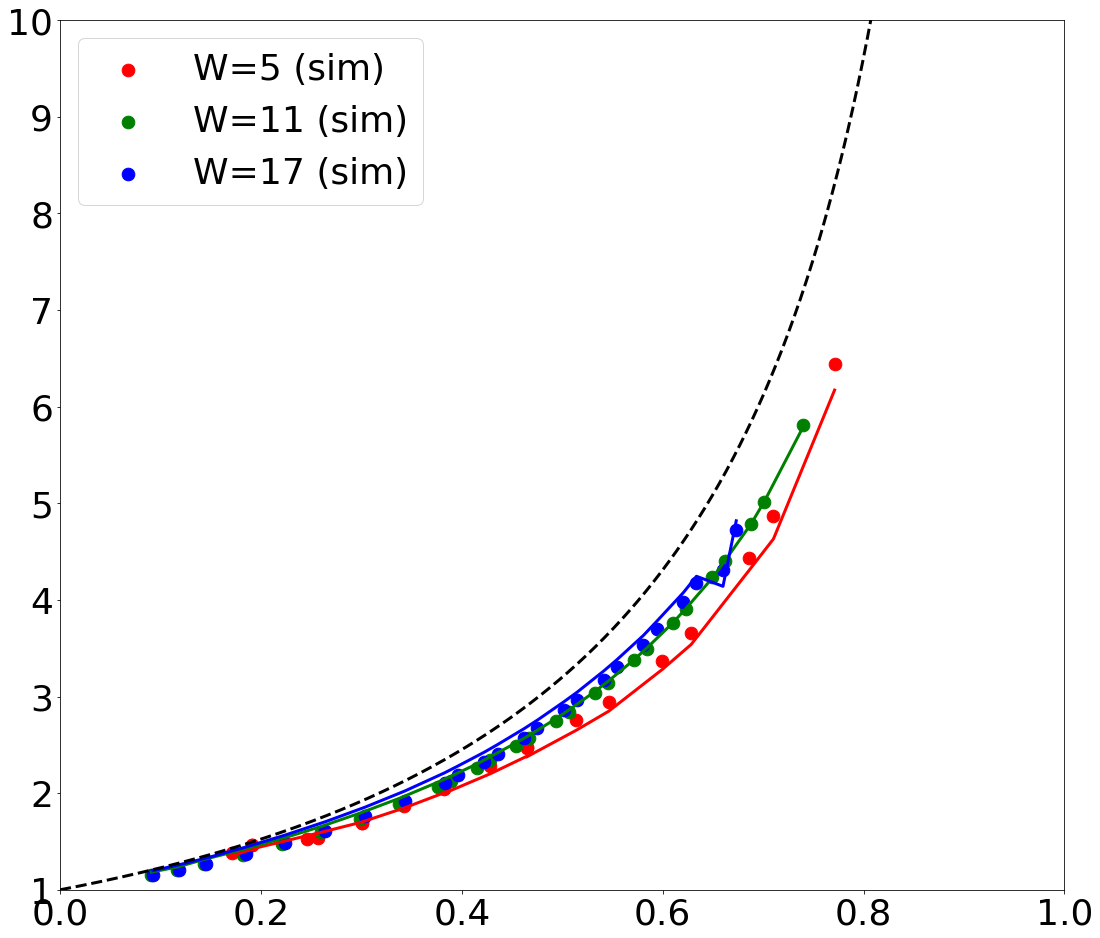}
\caption{\label{effective} Effective viscosity of a RBC suspension as a function of the global concentration of cells for several channel widths. The black dashed curve corresponds to the law (\ref{fit_law}) applied to the global concentration.}
\end{center}
\end{figure}

Finally as a test, we analyze the effective global viscosity $\eta_0$ of the confined suspension  ($\eta_0=1/f_0$), where $f_0=\int \bar f(y)dy/W$.
We compare $\eta_0$ predicted by the law (\ref{fit_law}) from the concentration profiles obtained in simulations to the effective viscosity measured directly, as shown in Fig. \ref{effective}.
We plot the effective viscosity as a function of the global concentration for different confinements.
We also add the result of the universal law (\ref{fit_law}) applied to the global concentration as a reference. 
Overall, we can see that the universal law (\ref{fit_law}) applied to the local concentration reproduces the numerical results quite well:
The  predicted values are much closer to the actual ones (curves and dots of the same color, respectively) than the result of applying eq. (\ref{fit_law}) to the global concentration (black dashed line). This means that the details of cell distribution plays an important role.

\paragraph*{Conclusion-}
We have presented here a universal microscopic law that governs the rheology of RBC suspensions.
The law allows us to make qualitative and often quantitative predictions about the concentration and fluidity profiles within the suspension. The method is clearly not limited to RBCs suspensions, but can also used for other systems (such as emulsions, for example).
Further steps can be taken to improve the predictive power of the proposed laws.
For the fluidity distribution, it is important to quantify the evolution of the local law with the capillary number, by including also the effect of plasma proteins (leading to RBCs aggreagtion).
Finally, how does the constitutive law translates to complex geometry (e.g. vascular network) is still unclear.
These questions constitute an important goal for future research.

We thank CNES (Centre National d'Etudes Spatiales) and the French-German University Programme "Living Fluids" (Grant CFDA-Q1-14) for the financial support. The simulations were performed on the Cactus cluster of the CIMENT infrastructure supported by the Rh\^one-Alpes region (GRANT CPER07\_13 CIRA).

\bibliographystyle{unsrt}
\bibliography{ref}
\end{document}


\preprint{APS/123-QED}

\title{Towards a universal law for blood flow}

\author{Alexander Farutin}
    \affiliation{Laboratoire Interdisciplinaire de Physique, Universit\'{e} Grenoble Alpes and CNRS, F-38000 Grenoble, France}
\author{Abdessamad Nait-Ouhra}
    \affiliation{Laboratoire Interdisciplinaire de Physique, Universit\'{e} Grenoble Alpes and CNRS, F-38000 Grenoble, France}
    \affiliation{LaMCScI, Faculty of Sciences, Mohammed V University of Rabat, 1014 Morocco}
    \author{Mehdi Abbasi}
    \affiliation{Laboratoire Interdisciplinaire de Physique, Universit\'{e} Grenoble Alpes and CNRS, F-38000 Grenoble, France}
    \author{Gopal Dixit}
    \affiliation{Laboratoire Interdisciplinaire de Physique, Universit\'{e} Grenoble Alpes and CNRS, F-38000 Grenoble, France}    \author{Hamid Ez-zahraouy}     \affiliation{LaMCScI, Faculty of Sciences, Mohammed V University of Rabat, 1014 Morocco}       
    \author{Othmane Aouane}
    \affiliation{Helmholtz Institut (HIERN)
Forschungszentrum J\"ulich GmbH
F\"urther Stra{\ss}e  248
90429 N\"urnberg, Germany}  \author{Jens Harting}
  \affiliation{Helmholtz Institut (HIERN)
Forschungszentrum J\"ulich GmbH
F\"urther Stra{\ss}e  248
90429 N\"urnberg, Germany}  \author{Chaouqi Misbah}
    \email{chaouqi.misbah@univ-grenoble-alpes.fr}
    \affiliation{Laboratoire Interdisciplinaire de Physique, Universit\'{e} Grenoble Alpes and CNRS, F-38000 Grenoble, France}

\date{\today}

\begin{abstract}
Here we show that under shear flow the stress is constant, whereas under Poiseuille flow it varies linearly with the lateral position, regardless of the constitutive law. We then  show data for weak shear-thinning, and provide more precision about  the nonlocal diffusion law.
. 

\end{abstract}

\pacs{Valid PACS appear here}
\maketitle


\paragraph*{The behavior of the stress in shear flow and under Poiseuille flow}
We focus here on 2D only. We will show that the only stress component $\sigma_{xy}$ is constant for a shear flow or varies linearly with the lateral coordinate for a Poiseuille flow. This proof is general and is based only on mechanical equilibrium (without any reference to a constitutive law). 

\begin{figure}
\begin{center}
\includegraphics[height=4 cm,width=0.99\columnwidth]{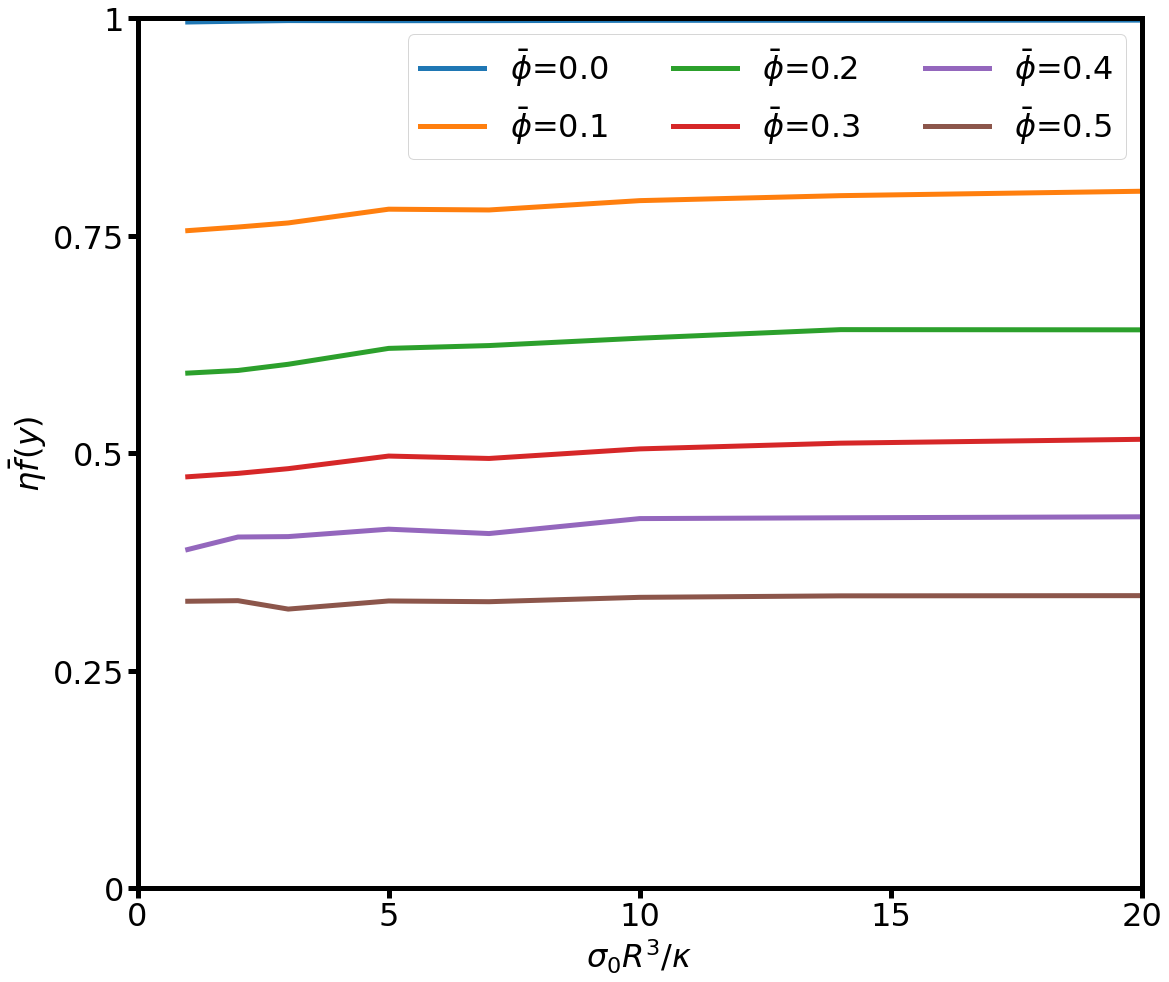}
\caption{\label{stress} Shear rate as a function of stress showing a weak shear-thinning.}
\end{center}
\end{figure}

Mechanical equilibrium  (we neglect inertia due to the Small Reynolds limit) implies: $div\boldsymbol{\sigma}=\mathbf{0}$, which leads to $\text{i) } \partial_{y} \sigma_{x y}=\partial_{x}\left(p-\sigma_{x x}\right) \text { and  ii) }  \partial_{x} \sigma_{x y}=\partial_{y}\left(p-\sigma_{y y}\right)$ where $p$ is the pressure, $x$ the flow direction and $y$ the spanwise direction. Differentiating the first equation with respect to $y$ and the second one with respect to $x$ and substracting the two resulting equations, we get $\partial_{y}^{2} \sigma_{x y}=\partial_{x}^{2} \sigma_{x y}-\partial_{y} \partial_{x}\left(\sigma_{x x}-\sigma_{y y}\right)=0$. Assuming that the flow is developed (meaning translational invariance along the flow direction $x$), we obtain $\partial_{y} \sigma_{x y}=C$, where $C$ is a constant. This implies that in general $\sigma_{x y}=Cy+D$ where $D$ is a constant. For a Poiseuille flow we set pressure equal to $P_1$ at $y=0$ and $P_2$ at $y=L$. Equating the total force $\sigma_{xy} n_x$ (where $n_x$ is the $x$ component of the normal (equal to $1$) we obtain finally $\sigma_{x y}=(\Delta P/L)y+P_1$, where $\Delta p=P_2-P_1$. In the absence of pressure difference (as for pure shear flow), we have $\sigma_{x y}=P_1\equiv \sigma_0$, a constant. 
\paragraph*{Shear-thinning}
The fluidity shows a weak shear-thinning behavior. Fig.\ref{stress} shows $\dot{\gamma}$ as a function of $\sigma$, the slope being the local fluidity.
\paragraph*{Nonlocal hydrodynamics diffusion}
The mass current is written as 
\begin{equation}
\label{Jr}
\bar J_{cy}^r=-D(y)\partial_y\bar\phi_c(y)+\bar\phi_c(y)\int\partial_y U(y-y')\bar\phi_c(y')dy',
\end{equation}
where $\bar\phi_c(y)$ represents the concentration of center of masses of cells; $D$ and $U$ are given below.  It is more practical to use this mass current instead of the local concentration itself $\bar\phi(y)$. The concentration profile is related to the center profile by a smoothing kernel $\bar\phi(y)=\int S(y,y')\bar\phi_c(y')dy'$, where $S(y-y')=\max{0,a_0-a_2(y-y')^2/R^2}$, $a_0=0.774643$, and $a_2=0.826386$ provides satisfactory results.

\begin{figure*}
\begin{center}
\includegraphics[width=0.65\columnwidth]{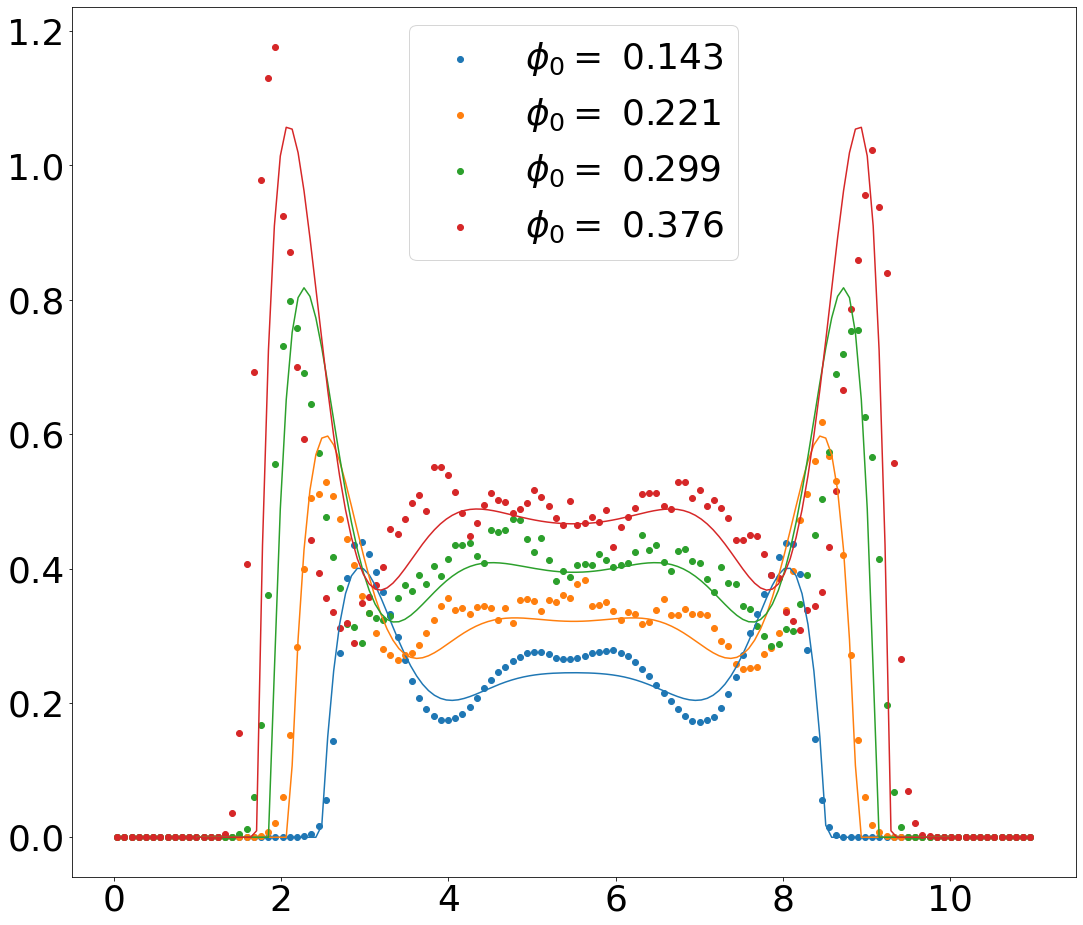}
\includegraphics[width=0.65\columnwidth]{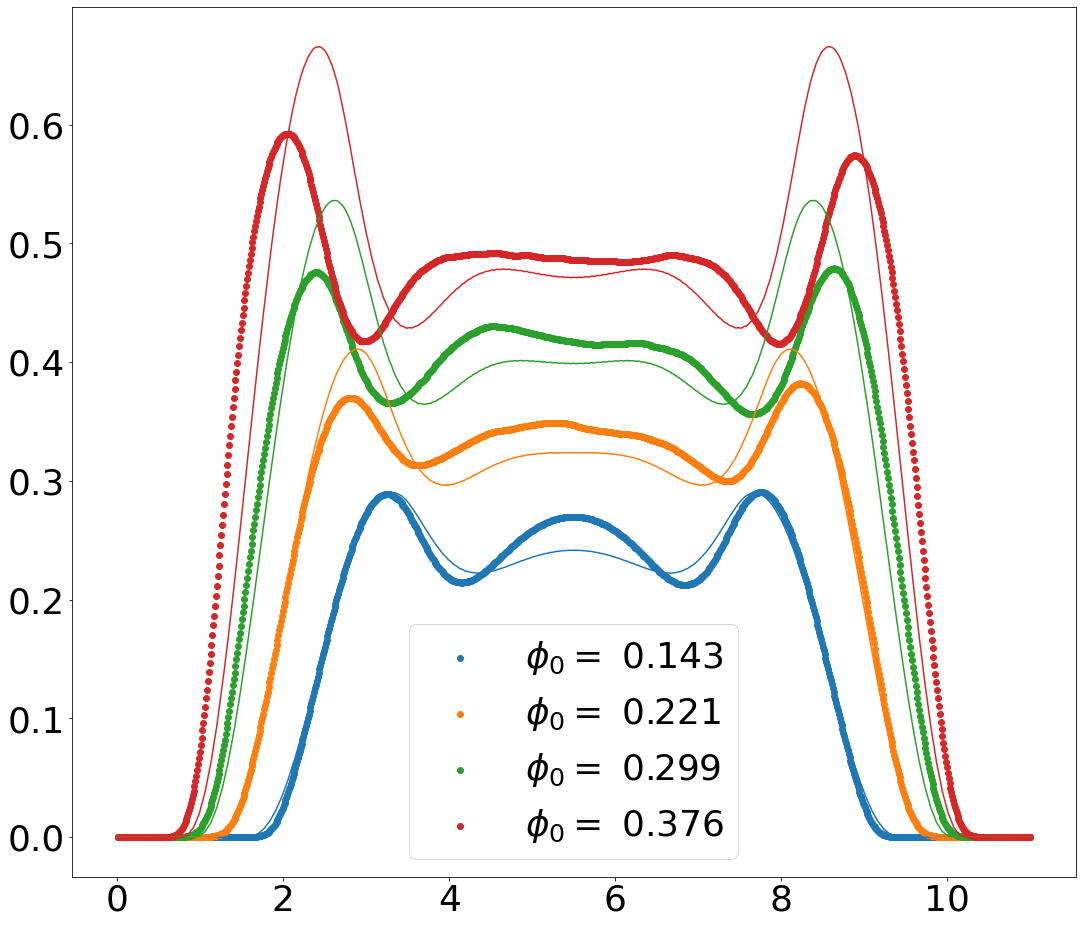}
\includegraphics[width=0.65\columnwidth]{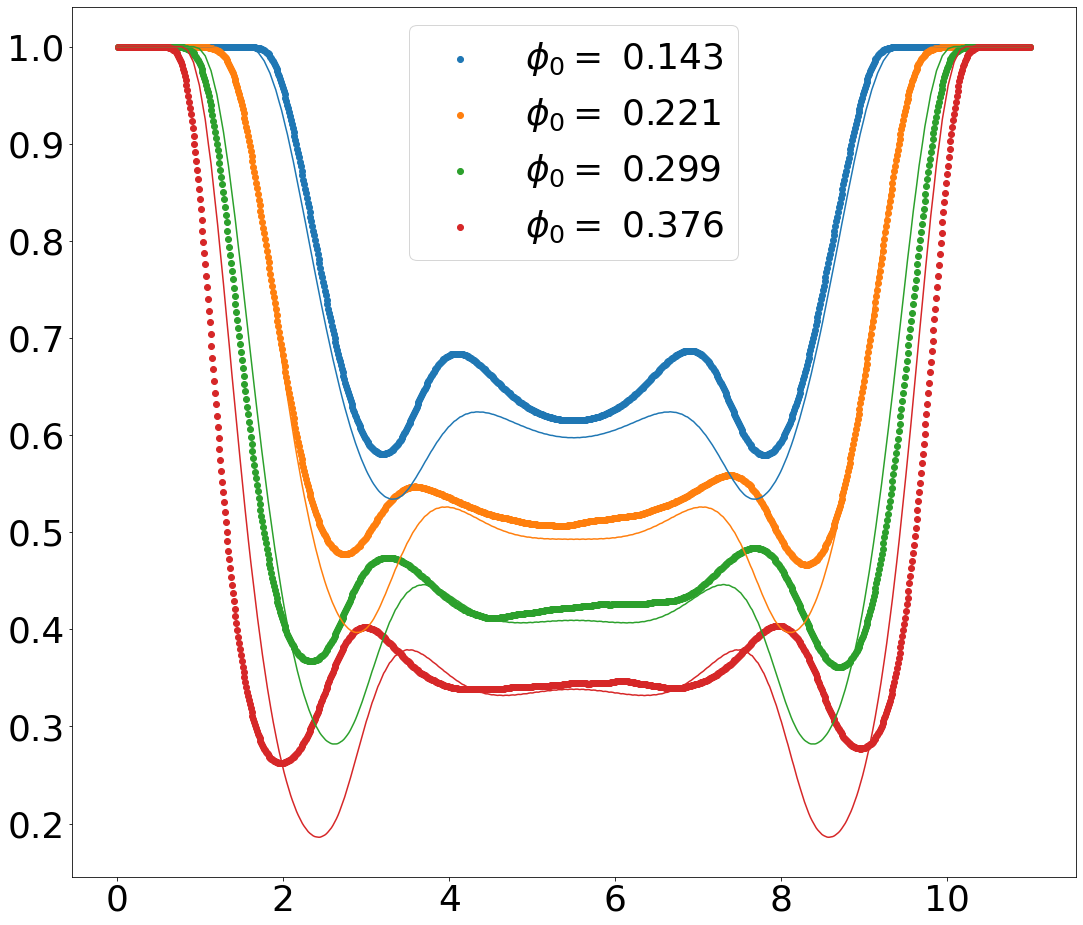}
\caption{\label{predicted}Predictions of the non-local migration model compared to simulation resutls. Left: Center of mass profiles. Center: Average local concentration. Right: Averge local fluidity.}
\end{center}
\end{figure*}

The local diffusion coefficient $D(y)$ depends on the concentration profile in the channel, but for simplicity, we have taken $D(y)=D_1\bar\phi_c(y)$, which is justified for the diffusive interaction that decays rapidly with the lateral separation of the vesicles. The deterministic repulsion potential $U$ defines the non-local diffusion in the suspension. We have taken $U(y-y')=D_2/(4R_0^2+(y-y')^2)$, which results in concentration profiles consistent with the numerical simulations (see Fig.\ref{predicted}).

\bibliographystyle{unsrt}
\bibliography{ref}